\documentclass[11pt,twoside,onecolumn]{article}
\usepackage[]{latexsym}
\usepackage{epsfig}
\usepackage{amsmath,amssymb}
\RequirePackage[dvipsnames,usenames]{color}
\newcommand{\be}{\begin{eqnarray}}
\newcommand{\ee}{\end{eqnarray}}

\title{\bf A simple method to generate exact physically acceptable anisotropic solutions in general relativity}
\author{J Ovalle$^{ab}$\thanks{jovalle@usb.ve}
$\,$ 
A Sotomayor$^{c}$\thanks{adrian.sotomayor@uantof.cl}
\\
\null
\\
$^a${\em Institute of Physics and Research Centre of Theoretical Physics and Astrophysics,}
\\
{\em Faculty of Philosophy and Science, Silesian University in Opava}
\\
{\em CZ-746 01 Opava, Czech Republic}
\\
$^b${\em Departamento de F\'{\i}sica, Universidad Sim\'on Bol\'ivar,}
\\
{\em AP 89000, Caracas 1080A, Venezuela}
\\
$^c${\em Departamento de Matem\'aticas, Universidad de Antofagasta}
\\
{\em  Antofagasta, Chile}
}
\begin{document}
\maketitle
\begin{abstract}
By using the gravitational decoupling through the minimal geometric deformation approach (MGD-decoupling), we show a simple and powerful approach to generate physically acceptable exact analytical solutions for anisotropic stellar distributions in general relativity. We find that some perfect fluid configurations could be incompatible with anisotropic effects produced by scalar fields.

\end{abstract}
%
%\keywords{Brane-world, Stars, Black holes}
%\pacs{04.50.+h, 04.70.-s, 04.70.Dy}
%
%
%
%
%
%
\newpage
\section{Introduction}
\setcounter{equation}{0}
The first simple, systematic and direct approach to
decoupling gravitational sources in general relativity (GR) was recently developed~\cite{MGD-decoupling} from the so-called
Minimal Geometric Deformation (MGD) approach (henceforth MGD-decoupling), an idea  originally proposed~\cite{jo1,jo2} in the context of the Randall-Sundrum
brane-world~\cite{lisa1,lisa2} and extended to investigate
new black hole solutions~\cite{MGDextended1,MGDextended2}
(For some earlier works, see for instance Refs.~\cite{exactbw,schwbw,jo6,jo8,jo9,jo10},
and for some recent applications Refs.~\cite{jo11,jo12,roldaoGL,rrplb,OCS,rr-glueball, rr-acustic,GLalysis,MGD bh,jcrs2017,camiloleon,glueball,milko,ernesto, tello,luciano,sharif1,sharif2,rperez,tello2,ernesto2,angel,milko2,ernesto3}).
\par
The MGD-decoupling has a number of ingredients that make it particularly useful in the search for new spherically symmetric solutions of Einstein's field equations. For instance, in order to find a solution to Einstein's equations with a complex spherically symmetric energy-momentum
tensor $T_{\mu\nu}$, we can split it into simpler components, say $T^{(i)}_{\mu\nu}$, namely
\be
\label{coupling}
T_{\mu\nu}=T^{(1)}_{\mu\nu}+T^{(2)}_{\mu\nu}+...
\ ,
\ee
and solve Einstein's equations for each one of these parts.
Hence, we will have as many solutions as are the contributions $T^{(i)}_{\mu\nu}$ in $T_{\mu\nu}$. 
Finally, by a straightforward combination of all these solutions, we will obtain the solution to the
Einstein equations associated with the original energy-momentum tensor $T_{\mu\nu}$. We stress that this method works as long as the sources do not exchange energy-momentum among them, namely
\be
\nabla_{\nu}T^{(1)\mu\nu}
=
\nabla_{\nu} T^{(2)\mu\nu}
=
\ldots
=
\nabla_{\nu} T^{(n)\mu\nu}
=
0
\ ,
\ee
\par
The approach described above represents a step forward 
in the search and analysis of solutions to Einstein's field equations, especially when we are studying situations beyond trivial
cases, such as the interior of stellar structures with gravitational sources more complex than the ideal perfect fluid~\cite{EM,singleton,sengupta,roman,marcelo1,Alvarez,carloni2017,ted,zdenek1,zdenek2,zdenek3,masud,sante2}. 
In this respect, the simplest practical application of the MGD-decoupling consists in extending known isotropic and physically acceptable interior solutions for spherically symmetric stellar systems into the anisotropic domain,
at the same time preserving physical acceptability, which represents a
highly non-trivial problem~\cite{Stephani}
(for obtaining anisotropic solutions in a generic way, see for instance
Ref.~\cite{lake3,luis1}).
\par
Finally, it is worth mentioning that the MGD introduces a deformation on the Schwarzschild solution that is characterized by a length scale $\ell$. This deformation and its consequences have been widely studied (see, for instance, Refs.~\cite{jo8,jo9,jo10,jo11,jo12,MGDBH,abdalla,hoff}). In the brane-world scenario, for instance, this length scale $\ell$ is proportional to the inverse of the brane tension $\sigma$ and therefore General Relativity is regained when $\sigma\to\infty$. We might be tempted to think that this length scale $\ell$ might imply, in some way, a breaking of the Lorentz symmetry. However, it is clear~\cite{jo10,rrplb} that the brane tension $\sigma$ is, at most, dependent on the universe temperature, that is a scalar. Indeed, it was found~\cite{abdalla,hoff}, by using a E\"otv\"os brane~\cite{laszlo}, that the length scale $\ell$ is proportional to the inverse of the temperature of the Universe and to the stellar distribution surface, which are both Lorentz invariant and general coordinate invariant.
\par
\par
The paper is organised as follows:
in Section~\ref{s2}, we briefly review the effective Einstein field equations
for a spherically symmetric and static distribution of matter with effective density
$\tilde{\rho}$, effective radial pressure $\tilde{p}_r$ and effective tangential pressure
$\tilde{p}_t$;
in Section~\ref{s3} we show two sets of equations generated when the MGD-decoupling is implemented on the effective Einstein equations displayed in Section~\ref{s2}; in Section~\ref{s4}, we provide a simple protocol with four basic steps, useful to extend perfect fluid solutions in the anisotropic domain; in Section~\ref{s5} we consider the Einstein-Klein-Gordon system and we find that some perfect fluid configurations could become unstable under the effects of scalar fields. We summarise our conclusions in Section~\ref{s6}.
\section{Einstein equations}
\label{s2}
\par
Let us start from the standard Einstein field equations
\begin{equation}
\label{corr2}
R_{\mu\nu}-\frac{1}{2}\,R\, g_{\mu\nu}
=
-k^2\,T_{\mu\nu}
\ ,
\end{equation}
where
\begin{equation}
\label{emt}
T_{\mu\nu}
=
(\rho +p)\,u_{\mu }\,u_{\nu }-p\,g_{\mu \nu }+\alpha\,\theta_{\mu\nu}
\ ,
\end{equation}
is the 4-dimensional energy-momentum tensor of a perfect fluid coupled to a generic  source $\theta_{\mu\nu}$, which in general produces anisotropies in self-gravitating systems. In addition to the 4-velocity field $u^\mu$, density $\rho$ and isotropic pressure $p$, some additional source like scalar, vector and tensor fields may be contained in the anisotropic sector represented by $\theta_{\mu\nu}$, with $\alpha$ as a coupling constant. As is well known, since the Einstein tensor is divergence free, the energy-momentum tensor~(\ref{emt})
must satisfy the conservation equation
\begin{equation}
\nabla_\nu\,T^{\mu\nu}=0
\ .
\label{dT0}
\end{equation}
In Schwarzschild-like coordinates, the spherically symmetric metric reads 
\begin{equation}
ds^{2}
=
e^{\nu (r)}\,dt^{2}-e^{\lambda (r)}\,dr^{2}
-r^{2}\left( d\theta^{2}+\sin ^{2}\theta \,d\phi ^{2}\right)
\ ,
\label{metric}
\end{equation}
where $\nu =\nu (r)$ and $\lambda =\lambda (r)$ are functions of the areal
radius $r$ only, ranging from $r=0$ (the star center) to some $r=R$ (the
star surface), with the 4-velocity given by 
$u^{\mu }=e^{-\nu /2}\,\delta _{0}^{\mu }$ for $0\le r\le R$. The Einstein equations are written as
\begin{eqnarray}
\label{ec1}
&&
-k^2
\left( \rho
+\alpha\,\theta_0^{\,0}
\right)
=-
\strut\displaystyle\frac 1{r^2}
+e^{-\lambda }\left( \frac1{r^2}-\frac{\lambda'}r\right)\ ,
\\
&&
\label{ec2}
-k^2
\strut\displaystyle
\left(-p+\alpha\,\theta_1^{\,1}\right)
=
-\frac 1{r^2}+e^{-\lambda }\left( \frac 1{r^2}+\frac{\nu'}r\right)\ ,
\\
&&
\label{ec3}
-k^2
\strut\displaystyle
\left(-p+\alpha\,\theta_2^{\,2}\right)
=
\frac 14e^{-\lambda }\left[ 2\,\nu''+\nu'^2-\lambda'\,\nu'
+2\,\frac{\nu'-\lambda'}r\right]
\ ,
\end{eqnarray}
while the conservation equation, which is a linear combination of Eqs.~(\ref{ec1})-(\ref{ec3}), yields
\begin{equation}
\label{con1}
-p'-\strut\displaystyle\frac{\nu'}{2}(\rho+p)+\alpha(\theta_1^{\,\,1})'-\strut\displaystyle\frac{\nu'}{2}\alpha(\theta_0^{\,\,0}-\theta_1^{\,\,1})-\frac{2\alpha}{r}(\theta_2^{\,\,2}-\theta_1^{\,\,1}) = 0
\ ,
\end{equation}
where $f'\equiv \partial_r f$.
We then note the perfect fluid equations are formally
recovered for $\alpha\to 0$.
\par
By simple inspection of the field equations~(\ref{ec1})-(\ref{ec3}), we
can identify an effective density 
\begin{equation}
\tilde{\rho}
=
\rho
+\alpha\,\theta_0^{\,0}
\ ,
\label{efecden}
\end{equation}
an effective isotropic pressure
\begin{equation}
\tilde{p}_{r}
=
p-\alpha\,\theta_1^{\,1}
\ ,
\label{efecprera}
\end{equation}
and an effective tangential pressure
\begin{equation}
\tilde{p}_{t}
=
p-\alpha\,\theta_2^{\,2}
\ .
\label{efecpretan}
\end{equation}
This clearly illustrates that the source $\theta_{\mu\nu}$ generates an anisotropy 
\begin{equation}
\label{anisotropy}
\Pi
\equiv
\tilde{p}_{t}-\tilde{p}_{r}
=
\alpha\,(\theta_1^{\,1}-\theta_2^{\,2})
\end{equation}
inside the stellar distribution.  
\par
The system Eqs.~(\ref{ec1})-(\ref{ec3}) contains five unknown functions, namely, 
three physical variables, the density $\tilde{\rho}(r)$, the radial pressure $\tilde{p}_r(r)$ and the tangential pressure $\tilde{p}_t(r)$, and two geometric functions: the temporal metric function $\nu(r)$ and the radial metric function $\lambda(r)$. Therefore these equations form an indefinite system. 

\section{Gravitational decoupling by MGD}
\label{s3}
\par
Let us implement the following ``geometric deformation" \cite{MGD-decoupling,jcrs2017} (Also see~\cite{OCS} and references therein for earlier works) on the radial metric component $g_{11}(r)$ by
\begin{eqnarray}
\label{expectg}
\mu(r)\rightarrow\,e^{-\lambda(r)}
=
\mu(r)+\alpha\,f^{*}(r)
\ ,
\end{eqnarray}
where $\alpha$ is a free parameter and $f^{*}(r)$ is the geometric deformation undergone by the radial metric component $\mu(r)$. Now let us plug the  decomposition in Eq.~(\ref{expectg}) in the Einstein equations~(\ref{ec1})-(\ref{ec3}). The system is thus   separated in two sets: (A) one having the standard Einstein field equations for a perfect fluid ($\alpha = 0$) of density $\rho$, pressure $p$, temporal metric component $g_{00}=e^{\nu}$ and radial metric component $g_{11}=-\mu^{-1}$ 
\begin{eqnarray}
\label{ec1pf}
&&
-k^2\rho
=-\frac{1}{r^2}+\frac{\mu}{r^2}+\frac{\mu'}{r}\ ,
\\
&&
\label{ec2pf}
-k^2
\left(-p\right)
=
-\frac 1{r^2}+\mu\left( \frac 1{r^2}+\frac{\nu'}r\right)\ ,
\\
&&
\label{ec3pf}
-k^2
\strut\displaystyle
\left(-p\right)
=
\frac{\mu}{4}\left(2\nu''+\nu'^2+\frac{2\nu'}{r}\right)+\frac{\mu'}{4}\left(\nu'+\frac{2}{r}\right)
\ ,
\end{eqnarray}
with the conservation equation yielding
\begin{equation}
\label{conpf}
p'+\strut\displaystyle\frac{\nu'}{2}(\rho+p) = 0
\ ,
\end{equation}
which is a linear combination of Eqs~(\ref{ec1pf})-(\ref{ec3pf}); and (B) one for the source $\theta_{\mu\nu}$, which reads
\begin{eqnarray}
\label{ec1d}
&&
-k^2\,\theta_0^{\,0}
=
\strut\displaystyle\frac{f^{*}}{r^2}
+\frac{f^{*'}}{r}\ ,
\\
&&
\label{ec2d}
-k^2
\strut\displaystyle
\,\theta_1^{\,1}
= f^{*}\left(\frac{1}{r^2}+\frac{\nu'}{r}\right)\ ,
\\
&&
\label{ec3d}
-k^2
\strut\displaystyle\,\theta_2^{\,2}
=\frac{f^{*}}{4}\left(2\nu''+\nu'^2+2\frac{\nu'}{r}\right)+\frac{f^{*'}}{4}\left(\nu'+\frac{2}{r}\right)
\ .
\end{eqnarray}
The conservation equation  $\nabla_\nu\,\theta^{\mu\nu}=0$ explicitly reads
\begin{equation}
\label{con1d}
(\theta_1^{\,\,1})'-\strut\displaystyle\frac{\nu'}{2}(\theta_0^{\,\,0}-\theta_1^{\,\,1})-\frac{2}{r}(\theta_2^{\,\,2}-\theta_1^{\,\,1}) = 0
\ ,
\end{equation}
which is a linear combination of Eqs.~(\ref{ec1d})-(\ref{ec3d}). 
Under these conditions, there is no exchange of energy-momentum between the perfect fluid and the source $\theta_{\mu\nu}$; their interaction is purely gravitational. 
\par
Next we develop a protocol, step by step, in a clear and straightforward way, hence  the reader will be able to generate, from any known perfect fluid solution for spherically symmetric self-gravitating systems, a new anisotropic solution.
\section{Protocol}
\label{s4}
\par
In order to find a solution $\{\tilde{\rho}(r), \tilde{p}_r(r), \tilde{p}_t(r), \nu(r), \lambda(r)\}$ to Einstein field equations in (\ref{ec1})-(\ref{ec3}), we will follow these simple steps:
\begin{itemize}
\item Pick up a perfect fluid solution $\{\rho, p, \nu, \mu\}$ to Einstein field equations shown in (\ref{ec1pf})-(\ref{ec3pf}). A (incomplete) list can be found in Ref.~\cite{lake1} (See also \cite{lake2,visser2005}).  

\item Impose the ``mimic constraint" defined by 
\[
\theta_1^{\,1}(r)=p(r)\ .
\]
Hence the geometric deformation $f^{*}$ is automatically determined by Eq.~(\ref{ec2d}). At this stage we have both $f^{*}(r)$ and $\nu(r)$ determined. 

\item Use the ``MGD-transformation" in (\ref{expectg}) to find the metric function $\lambda(r)$. At this stage we have both metric components in (\ref{metric}), namely, $\nu(r)$ and $\lambda(r)$.

\item Use the metric functions $\{\nu(r), \lambda(r)\}$ in the field equations (\ref{ec1})-(\ref{ec3}) to find the physical variables $\{\tilde{\rho}(r), \tilde{p}_r(r), \tilde{p}_t(r)\}$ defined in (\ref{efecden})-(\ref{efecpretan}).
\par
To see a specific example, the reader may follow the reference Ref.~\cite{jcrs2017}, where the construction of the anisotropic version of the well known Tolman IV solution for perfect fluids~\cite{Tolman} is developed in detail. This despite the present recipe had not yet been developed.
\end{itemize}
\section{Einstein-Klein-Gordon}
\label{s5}
\par
So far we have considered a generic anisotropic source $\theta_{\mu\nu}$ which produces anisotropic consequences on any perfect fluid solution $\{\tilde{\rho}=\rho;\,\tilde{p}_r=p;\,\tilde{p}_t=p\}$, as is clearly shown through Eqs.~\eqref{ec1}-\eqref{anisotropy}. Now a natural question arises. What happens if the generic source $\theta_{\mu\nu}$ has a specific form associated with a known anisotropic system? Let us say, a scalar field $\psi$ described by the Klein-Gordon equation? We know many alternatives to general relativity containing scalar fields, and these scalar fields have a direct impact on self-gravitating systems. 
\par
Let us recall the action for a scalar field $\psi$ with a potential $V(\psi)$ is
given by
\begin{equation}
\label{KGaction}
S
=
\int\left[\frac{1}{2}\,\nabla_\mu\psi\nabla^{\mu}\psi-V(\psi)\right]
\sqrt{-g}\,d^{4}\,x
\ ,
\end{equation}
which yields the Klein-Gordon equation
\begin{equation}
\label{KGeqmotion}
\nabla_\mu\nabla^{\mu}\psi+\frac{dV}{d\psi}
=
0
\ .
\end{equation}
The energy-momentum tensor $\theta_{\mu\nu}$ of the scalar field $\psi$ is given by
\begin{equation}
\label{KGem2}
\theta_{\mu\nu}=
\nabla_\mu\psi\,\nabla_\nu\psi
-\left(\frac{1}{2}\,\nabla_\alpha\psi\,\nabla^\alpha\psi-V\right)
g_{\mu\nu}
\ .
\end{equation}
Therefore, when a perfect fluid solution $\{\tilde{\rho}=\rho;\,\tilde{p}_r=p;\,\tilde{p}_t=p\}$ is considered along with the scalar field $\psi$, namely, the Einstein-Klein-Gordon system for the interior of a self-gravitating distribution, the MGD-decoupling yields the deformed metric~\eqref{expectg}, whose deformation $f^{*}$ satisfies the set~\eqref{ec1d}-\eqref{ec3d}, which in this case reduces to
\begin{eqnarray}
\label{kg1d}
&&
k^2
\left[\frac{1}{2}\,e^{-\lambda}\,{\psi'}^2+V\right]
=-\frac{f^{*}}{r^2} 
-\frac{f^{*'}}{r}\ ,
\\
&&
\label{kg2d}
k^2
\left[-\frac{1}{2}\,e^{-\lambda}\,{\psi'}^2+V\right]
=
-f^{*}\left(\frac{1}{r^2}+\frac{\nu'}{r}\right)\ ,
\\
&&
\label{kg3d}
k^2
\left[\frac{1}{2}\,e^{-\lambda}\,{\psi'}^2+V\right]
=
-\frac{f^{*}}{4}\left(2\,\nu''+\nu'^2+2\frac{\nu'}{r}\right)
-\frac{f^{*'}}{4}\left(\nu'+\frac{2}{r}\right)
\ ,
\end{eqnarray}
and the conservation equation~\eqref{con1d} reads
\begin{equation}
\label{conKG2}
\psi''+\left[\frac{2}{r}+\frac{1}{2}\left(\nu'-\lambda'\right)\right]\psi'
=
e^{\lambda}\,\frac{dV}{d\psi}
\ ,
\end{equation}
which is nothing but the explicit form of the Klein-Gordon equation~\eqref{KGeqmotion}.
\par
Since the temporal metric component in~\eqref{metric} is not deformed under the MGD, 
the system~\eqref{kg1d}-\eqref{conKG2} contains three unknown functions $\{f^{*},\,\psi,\,V\}$ to be determined by three independent equations in Eqs.~\eqref{kg1d}-\eqref{conKG2}. Hence, contrary to the second step in the protocol of Section~\ref{s4}, no further restriction is necessary. The above is very significant for two reasons. First, we see the protocol works very well as long as the anisotropic source $\theta_{\mu\nu}$ remains generic~\cite{MGD-decoupling,jcrs2017,camiloleon,milko,tello,sharif2,tello2}. However, if we consider a specific gravitational source $\theta_{\mu\nu}$ which does not have enough degrees of freedom, the protocol could fail. Secondly, in the case of the scalar field $\psi$,  every perfect fluid solution $\{\rho,\,p,\,\nu,\,\mu\}$ to the system~\eqref{ec1pf}-\eqref{conpf} will have a specific scalar configuration $\{\psi,\,V\}$ associated, which will produce a specific anisotropy on the perfect fluid through the deformation $f^{*}$. Since there are many perfect fluid solutions, this means that we have a scheme rich in possibilities to be exploited. However, and this is fair to say, as soon as we choose a perfect fluid solution $\{\rho,\,p,\,\nu,\,\mu\}$, the system~\eqref{kg1d}-\eqref{conKG2} becomes a very restricted one, which could lead to a potential $V(\psi)$ with no physical meaning. We conclude some perfect fluid configurations could become unstable under the effects of scalar fields. This is a point that deserves to be investigated. However this is beyond the objective of this article.

\section{Conclusions}
\label{s6}
\par
By using the MGD-decoupling, we show in detail a simple and powerful method to extend all known isotropic and physically acceptable interior solutions for spherically symmetric stellar systems into the anisotropic domain. A key point concerning the extension of the physical acceptability, which indeed is a critical and non-trivial issue, is the mimic constraint $\theta_1^{\,1}(r)=p(r)$. This condition ensures that the generated anisotropic solution inherits the physical acceptability of the seed isotropic solution. However, when a specific form for the generic gravitational source $\theta_{\mu\nu}$ is considered, like the energy-momentum tensor associated with a scalar field $\psi$, the approach could fail. This clearly show that any specific gravitational source $\theta_{\mu\nu}$ should have enough degrees of freedom to successfully implement the protocol described in Section~\ref{s4}. In this respect, the system~\eqref{kg1d}-\eqref{conKG2} may be used to find the scalar configuration $\{\psi,\,V\}$ associated with a specific perfect fluid solution. Hence we could elucidate whether this configuration is stable or not.   
\par
We would like to end by mentioning that the source of the anisotropy, encoded in the generic energy-momentum tensor  $\theta_{\mu\nu}$, could have any origin. It could even be associated with modifications to general relativity.
\subsubsection*{Acknowledgements}
\par

J.O.~is supported by the Albert Einstein Centre for Gravitation and Astrophysics financed by the Czech Science Agency Grant No.14-37086G. A.S.~ is partially supported by Project Fondecyt 1161192, Chile. This work was supported by MINEDUC-UA project, code ANT 1755

\end{document}